\documentclass[10pt]{iopart}
\usepackage{iopams}
\usepackage[dvips]{graphicx}
\usepackage[scriptsize,bf]{caption}
\usepackage{xspace}
\eqnobysec
\renewcommand{\br}[1]{\left(#1\right)}

\renewcommand{\vec}[1]{\boldsymbol{#1}}
\newcommand{\abs}[1]{\left|#1\right|}

\newcommand{\csp}{$\mathbb{R}^3\times \mathbb{S}^2$\xspace}
\newcommand{\dv}[1]{\dot{\vec{{#1}}}}
\newcommand{\dd}[1]{\dot{#1}\dot{#1}}
\newcommand{\ddv}[1]{\ddot{\vec{{#1}}}}

\newcommand{\dsv}[1]{{\vec{{#1}}}'}
\newcommand{\eps}{\epsilon}

\newcommand{\eqref}[1]{\eref{#1}}
\newcommand{\frr}{fundamental relativistic rotator\xspace}

\newcommand{\sgn}{\mathrm{sgn}}
\newcommand{\sq}[1]{\left[#1\right]}

\newcommand{\ud}[1]{\mathrm{d}#1}
\newsavebox{\oldtext}

\newcommand{\myfootnote}[1]{$^\arabic{footnote}$\footnotetext{$^\arabic{footnote}${#1}}\addtocounter{footnote}{1}}

\begin{document}

\title[\underline{{\L}ukasz Bratek, \ \ \ \ \ \ \ \ \ \ \ \   False constraints. A toy model\ \dots\dots \ \ \ (preprint)} ]{False constraints. A toy model for studying dynamical systems with degenerate Hessian form.
}
\author{{\L}ukasz Bratek}
\address{Henryk Niewodnicza{\'n}ski Institute of Nuclear Physics, \\
Polish Academy of Sciences, Radzikowskego 152, PL-31342 Krak{\'o}w, Poland}
\ead{lukasz.bratek@ifj.edu.pl}
\begin{abstract}\\ This paper studies various aspects of the motion of relativistic rotators, both in the presence and absence of external fields, using a toy model which, in a sense, can be regarded as a non-relativistic limit of the rotators.
In a simpler setup, this enables one  to gain an insight into the principal difference between
mechanical systems with singular and non-singular Hessian, whilst avoiding
the complications resulting from the more intricate form of the equations of motion in the fully relativistic regime.
In particular, one can comprehend
the apparent contradiction between Hessian singularity and simultaneous occurrence of unique solutions for motion of the \frr minimally coupled to the electromagnetic field.
With the aid of the toy model the author supports and illustrates his thesis put forward elsewhere that the Hessian singularity is a defect
that makes physically unviable some geometric models of spinning particles considered in the literature.
\end{abstract}
\pacs{03.30.+p, 45.50.-j}
\medskip
\textbf{The definitive version is available at \\ \texttt{http://iopscience.iop.org/1751-8121/43/46/465206}}
\hrule

\section{Introduction}
By definition, a well-behaved mechanical system has its motion uniquely determinable from the stationary action principle and initial conditions. The Euler-Lagrange equations provide only a necessary condition for an
extremum of an action functional with respect to continuous comparison functions. A particular solution  can furnish an actual extremum only if it satisfies certain additional conditions. In particular, for a well-behaved dynamical system
it is assumed that its Lagrangian $L(q,\dot{q})$ should be a convex function of generalized velocities $\dot{q}$ associated with
the dynamical degrees of freedom $q$ \cite{bib:arnold}. The degrees of freedom $q$ are called dynamical to distinguish them from other gauge or auxiliary degrees of freedom, such as Lagrange multipliers or arbitrary world-line parameter, that are irrelevant to the dynamics. Non-dynamical degrees of freedom are assumed to have been already eliminated or fixed. In that case, convexity of the Lagrangian means in practice that the \textit{Hessian matrix} $\mathcal{H}$ with elements  $$\mathcal{H}_{ij}\equiv\frac{\partial^2 L(q,\dot{q})}{\partial \dot{q}^i\partial \dot{q}^{j}},\qquad i,j=1,2,\dots, N,$$ should be positive definite. Here, $N$ stands for the number of dynamical degrees of freedom.
This necessary condition  for a minimum of the action functional  is known as (strong) Legendre's necessary criterion \cite{bib:hilbert}.
 Positive definiteness requires, in particular, that the \textit{Hessian determinant} should be nonzero
 $$\left|{\frac{\partial^2 L(q,\dot{q})}{\partial \dot{q}^i\partial \dot{q}^{j}}}\right|\ne0,$$ in the appropriate domain.  The latter condition is necessary for  invertibility of a map from the momenta $p_i=\partial_{\dot{q}^i}L(q,\dot{q})$ to the velocities $\dot{q}^i$, that is, for equivalence  of Euler-Lagrange's and Hamilton's
equations of motion. The Legendre necessary criterion is not sufficient, there are more restrictive conditions for an actual minimum of the action integral. However, it is sufficient for our needs because
its violation suffices to ascertain that a mechanical system is defective.
 \textit{A mechanical system is called defective when the Hessian determinant is identically zero.}
 One of the purposes of this work is to support the statement that \textit{by its very nature, a defective dynamical system should not be considered as physical.}

In a recent paper \cite{bib:bratek_1}  it was demonstrated that the geometrical model of a spinning particle
suggested by Kuzenko, Lyakhovich and Segal
\cite{bib:segal_1}, and independently    rediscovered later in quite a different context  as a fundamental relativistic rotator
by Staruszkiewicz
\cite{bib:astar1}, is defective.  This model is described by five dynamical degrees of freedom, three associated with the position in physical space ($\thicksim\mathbb{R}^3$) and two additional degrees of freedom associated with the direction in space ($\thicksim\mathbb{S}^2$). The configuration space is thus identical to \csp, like for all other relativistic rotators (the definition of a relativistic rotator and the most general form of its Lagrangian can be found in \cite{bib:astar1}). It was found later \cite{bib:bratek_4} that the equations governing the evolution of \frr on \csp are linearly dependent.

The \frr is the only relativistic rotator whose Casimir invariants of the Poincar\'{e} group are fixed parameters, independently of the state of motion (to be more precise, there is a sign ambiguity leading to two such systems \cite{bib:bratek_1,bib:schaefer}).   This exceptional property of a mechanical system was originally used as a basic principle to fix a unique action functional \cite{bib:segal_1}. Many years later,  essentially the same principle was arrived at in \cite{bib:astar1} based on Wigner's idea of classification of relativistic quantum mechanical systems \cite{bib:wigner}. Staruszkiewicz  pointed out that quantum irreducibility, being simple algebraic notion, has its classical counterpart  (unlike unitarity), which, in the context of relativistic classical mechanical systems means that both Casimir invariants of
the Poincar\'{e} group should be parameters with fixed numerical values (that is, be independent of the state of
motion). Classical dynamical systems with this property are called \textit{fundamental} \cite{bib:astar1} and the conditions that mass and spin should be fixed parameters are called \textit{fundamental conditions}.
Owing to this fundamental property, the \frr was hoped to provide a particularly interesting dynamical system that could also be used to study some important and not well-understood theoretical issues \cite{bib:astar1}.
In this context, the unexpected Hessian singularity of the \frr discovered in \cite{bib:bratek_1} appears indeed disappointing.

The motion of the \frr can be regarded as a composition of two basic motions.
The first, inertial one, of the center of momentum frame (determined by the conserved momentum four-vector) and the second one, a circular motion in this frame about a circle of fixed radius (lying in a spatial plane orthogonal to the conserved Pauli-Luba\'{n}ski spin-pseudovector). This is what might be expected
for a classical spinning particle. The frequency of rotation, however, turned out to remain completely indefinite.  This frequency may be an arbitrary function of the proper time in the center of momentum  frame. In effect, there are possible infinitely many distinct motions satisfying identical initial conditions! Expressed in a covariant form, a parametric description of the most general solution to the equations of motion of the \frr can be found in \cite{bib:bratek_1}. Observe, that the motion of other (non-fundamental) relativistic rotators is definite, their Hessian determinants are nonzero.

In the conclusion to  \cite{bib:bratek_1} it was pointed out that the Hessian singularity of the \frr could be removed by supplementing its Lagrangian with a \textit{suitable} interaction term. Almost exactly at the same time, a preprint of another paper appeared in which an electrically charged version of the \frr  was suggested \cite{bib:schaefer}. The original Lagrangian was there supplemented with a minimal interaction term with the electromagnetic
 field. This tacitly assumes that the rotator can be treated as a structureless point-like particle. But this form of interaction is not suitable to remove the Hessian singularity. This becomes evident almost immediately by  writing the minimal interaction term in the usual vector notation   $L_{INT}=e\br{\frac{\dot{\vec{x}}}{c}\vec{A}(\vec{x},t)- \Phi(\vec{x},t)}$. This interaction is linear in the velocities, thus cannot contribute to the Hessian (compare with the calculation of the Hessian determinant carried out in \cite{bib:bratek_1}). For that reason one has to conclude that the dynamical system suggested in \cite{bib:schaefer} is defective, similarly as the (electrically neutral) \frr. This conclusion, however, may appear to contradict the statement made in \cite{bib:schaefer} that "in realistic situations, when a constant external field is
present, the frequency of rotation is as a rule constant and fixed in value". This statement was based on   a particular motion in the uniform magnetic field studied therein. Could this statement be true independently of the form of the electromagnetic field?
  It would be rather surprising to find out that the motion of the system was made unique merely because the external field has been switched on (irrespective of its nature and actual properties). Imagine only a situation in which an electrically charged \frr, having extremely large mass and spin, initially moving through empty space and rotating with indefinite frequency,  finally enters a zone of extremely weak electromagnetic wave (say, produced by a distant cell phone) and its motion immediately becomes unique and instantaneously correlated with arbitrary changes in this field. Absurdity of this situation reveals that something is indeed wrong with that model. The motion of the rotator must be constrained
   by the following condition $\vec{n}\circ\br{\frac{\dv{x}}{c}\times\vec{H}+\vec{E}}
=\frac{\dv{x}}{c}\circ\vec{E}$ found in \cite{bib:schaefer}. Here, $\vec{E}$ and $\vec{H}$ are vectors of the electric and magnetic field, respectively, $\vec{n}$ is the unit direction of the rotator and $\dv{x}$ is the velocity of the position vector. This constraint can be regarded as a consistency condition of the equations of motion and fundamental conditions. Note that this constraint does not depend on the electric charge, thus, it is independent of the "strength" with which the rotator couples to the electromagnetic field, and it also does not depend on the intensity of the electromagnetic field  --  the constraint can be multiplied by an arbitrary scalar function. This suggest that the constraint is not of a dynamical origin and must be somehow connected with the form of the Lagrangian of the \frr. In ref \cite{bib:bratek_4} it was shown that this is indeed the case, this constraint appears because the Hessian determinant vanishes, and for rotators with non-singular Hessian (whose motion is always definite)  no constraint appears.

As something of an aside, it is worth mentioning that following paper \cite{bib:bratek_1}, it was established that two other Lagrangians in Minkowski spacetime, constructed based on the same basic principle of fixed mass and spin, also define
 defective dynamical systems \cite{bib:bratek_2}. One of the Lagrangians turned out to have been already considered in \cite{bib:segal_2}.
 One might wonder whether there is any direct connection between
  fundamental conditions and Hessian singularity. This important theoretical issue still remains unanswered.
  Meanwhile,  more general dynamical systems consisting of a worldline and a single spinor are being investigated  in the hope of finding a well-behaved fundamental dynamical system with non-singular Hessian. A particular example of such a system can be regarded as a relativistic counterpart of axisymmetric top (with the spinor's phase interpreted as an angle of rotation about the spinor's direction) \cite{bib:bratek_3}.

Let us come back to the main subject.
   Among other reasons for the current paper is to elucidate the apparent contradiction unfolded above, by studying analogous behavior in a simpler model.
  To this end we use a toy model which, on the one hand, can be treated as distinct and completely unrelated to relativistic rotators, and, on the other hand, owing to its construction,  could be regarded as a non-relativistic limit of these rotators. We shall see in a simpler setup the basic difference between the motion of well-behaved dynamical systems (with non-singular Hessian) and the motion of defective dynamical systems (with singular Hessian), both in the presence and absence of  external fields. To make the toy model as much similar to that suggested in \cite{bib:schaefer} as possible, we supplement the free Lagrangian with the ordinary additive interaction term with the electromagnetic field.  This work is written in the context of a concrete model, however, the presented way of reasoning is quite universal and could be equally well applied to study other dynamical systems. In particular, this paper parallels an analogous analysis  carried out in \cite{bib:bratek_4} for electrically charged \frr in a fully relativistic case.

Before entering into the details of all this, it is appropriate to make some remark regarding the limit of relativistic rotators we consider here. It should be clear that  one can always consider the limit of slow motions. It is customary to do this merely by cutting off certain formal expansions of Lagrangians or equations of motion. However, one should be warned that such procedure is not justified unless it is proved that a particular "cutoff" indeed leads to solutions that approximate the original situation.
We shall not enter into this important and often neglected issue, since the aim of this paper is to illustrate what a defective dynamical system is, and to substantiate our thesis that such systems should not be considered as physical.
   The clue behind the toy model is that the Hessian singularity is a feature of the fundamental rotator that is unrelated to the speed values considered.
    Moreover, the non-relativistic limit simply makes the given examples easier to tackle with, thus more pedagogical and transparent,  leaving apart all the
  problems connected with solving more involved equations. Instead of having a family of relativistic rotators enumerated by a real function of single variable (leading to equations that in general cannot be solved in analytical way), we construct a family of non-relativistic rotators enumerated by a single continuous parameter and with equations of motion that can be solved.

 \section{A class of non-relativistic rotators in free motion}
Let us first consider the class of relativistic rotators defined in \cite{bib:astar1}  by the following Hamilton's action
\begin{equation}\label{eq:action_relativ_rotators}S=-m c\int\ud{\tau} \sqrt{\dd{x}}f\br{-\ell^2\frac{\dd{k}}{\br{k\dot{x}}^2}}, \qquad kk=0.\end{equation}
Here, $f$ is a sufficiently smooth function of a single variable and $k$ is a null direction.
In a map in which $x^{\mu}=\{c\,t,\vec{x}\}$, $k^{\mu}=\{1,\vec{n}\}$, where $\vec{n}$ is a unit vector parameterized by the spherical angles $\theta$ and $\phi$, the action reads
$$S=-mc^2\int\ud{t} \sqrt{1-\frac{\dv{x}^2}{c^2}}\,\mathcal{F}\!\br{\frac{\frac{\ell}{c}
\abs{\dv{n}}}{{1-\frac{\vec{n}\dv{x}}{c}}}}.
$$
Obviously, owing to the symmetries of the original action, this map is admissible. The arbitrary worldline parameter $\tau$ has been identified with the inertial time, $\tau\equiv t$.  This map also fixes the arbitrary scale of the null direction. Thereby, only five degrees of freedom are left in the Lagrangian which are  dynamical, and all the other, uninteresting, non-dynamical degrees of freedom have been eliminated.
The function $\mathcal{F}$ is trivially related to $f$. One may assume that $\mathcal{F}(0)=1$ and $\abs{\mathcal{F}'(0)}=1$ on account of the presence of two arbitrary (positive) dimensional parameters $m$ and $\ell$.

In the non-relativistic limit, $\abs{\dv{x}}\ll{}c$ and $\ell\abs{\dv{n}}\ll{}c$, the action integral \eqref{eq:action_relativ_rotators} defines a mechanical system with the following Lagrangian (the constant $-m c^2$ has been omitted):
\begin{equation}\label{eq:lagrangian}L=\frac{1}{2}m\dot{\vec{x}}^2+\frac{1}{2}a_2m\ell^2\dot{\vec{n}}^2- a_1 m \ell c \abs{\dot{\vec{n}}}\br{1+ \frac{\vec{n}\dot{\vec{x}}}{c}}.\end{equation}
This Lagrangian was derived by neglecting the higher order terms in a formal Taylor series expansion in $c^{-1}$ of the original Lagrangian.
The dimensionless parameters $a_1$ and $a_2$ are simply related to the function $\mathcal{F}$: $
a_1=\mathcal{F}'(0)$ and $ a_2=-\mathcal{F}''(0)$. One can also use an auxiliary  parameter $\beta=a_1-a_2/a_1$, to enumerate the above family of non-relativistic rotators, similarly as the function $f$ enumerates the family of relativistic rotators.

\subsection{Solutions of the equations of motion}
The procedure of determining solutions can be made coordinate independent by
means of finding the extremals of the action functional with the subsidiary condition $\vec{n}^2=1$.
This can be achieved by supplementing the
Lagrangian \eqref{eq:lagrangian} with an additive term $L_{\Lambda}=\Lambda\br{1-\vec{n}^2}$
with $\Lambda$ being yet an unknown function of time, disregarding at the same time the subsidiary condition
 which will reappear  as one of equations \cite{bib:hilbert}.
The function $\Lambda$ plays the role of an additional degree of freedom, but it is not dynamical because of identical vanishing of the momentum canonically conjugated to $\Lambda$. The function $\Lambda$ may be eliminated from the equations of motion and, had $\vec{x}$ and the unit vector $\vec{n}$ have been found, also uniquely specified.

The momenta canonically conjugated to $\dv{x}$ and $\dv{n}$ are
\begin{equation}\fl\label{eq:momenta}\vec{p}=m\dv{x}-a_1 m\ell \abs{\dv{n}}\vec{n}, \qquad \vec{\pi}= m\ell c \br{a_2\frac{\ell}{c}\abs{\dv{n}}-a_1 \br{1+\frac{\vec{n}\vec{\dv{x}}}{c}}}\frac{\dv{n}}{\abs{\dv{n}}}.\end{equation}
They are formally identical to the momenta of the original system for which $\vec{n}^2=1$. If $\vec{n}^2\equiv1$, then $\vec{\pi}\vec{n}\equiv0$, whereas, in general, $\vec{\pi}$ has also the longitudinal component along vector $\vec{n}$.
Since $\partial_{\vec{x}}L=0$, the momentum $\vec{p}$ is conserved. Let $\vec{p}_o$ denote the constant value of $\vec{p}$, then $$\vec{x}(t)=\frac{\vec{p}_o}{m}t+a_1\ell\int\abs{\dv{n}}\vec{n}\,\ud{t}.$$
The first term describes the uniform motion of the center of mass. The second term describes the intrinsic motion of the system as perceived in the center of mass frame. The motion of vector $\vec{n}$ is determined from the other part of Euler-Lagrange equations
$$\dv{\pi}=\frac{\partial{L}}{\partial\vec{n}}+\frac{\partial{L_{\Lambda}}}{\partial\vec{n}}, \qquad 0=\frac{\partial{L_{\Lambda}}}{\partial\Lambda}.$$
On multiplying the first equation by $\vec{n}$, we obtain an explicit expression for $\Lambda$
$$\Lambda=\frac{1}{2}\br{\vec{n}\frac{\partial{L}}{\partial\vec{n}}-
\vec{n}\dv{\pi}},\qquad \abs{\vec{n}}=1.$$
This leads us to the following equation for $\vec{n}$:
\begin{equation}\label{eq:eq_for_n_general}\vec{P}_{\vec{n}}\sq{\dv{\pi}-\frac{\partial{L}}{\partial\vec{n}}}=0,\qquad \abs{\vec{n}}=1.\end{equation}
Here, $\vec{P}_{\vec{n}}$ is a projection operator onto the subspace orthogonal to $\vec{n}$
$$\vec{P}_{\vec{n}}=\vec{1}-{\vec{n}\otimes\vec{n}},\qquad \vec{P}_{\vec{n}}[\vec{w}]=\vec{w}-\br{\vec{n}\vec{w}}{\vec{n}},\qquad \abs{\vec{n}}=1.$$
From now on we shall be assuming that $\abs{\vec{n}}\equiv1$.
When $\abs{\dv{n}}\ne0$, instead of the time variable $t$, one can introduce the natural, arc length parameter $s$ on the unit sphere ${\vec{n}}^2=1$.  Differentiation with respect to $s$ will be denoted by a 'prime' sign. On expressing $\dv{x}$ in terms of the vector $\vec{p}_o$, and
taking advantage of the identity
  $$\vec{p}_o=\br{\vec{p}_o\vec{n}}\vec{n}+\br{\vec{p}_o\dsv{n}}\dsv{n}+
  \br{\vec{p}_o\br{\vec{n}\times\dsv{n}}}\vec{n}\times\dsv{n},$$
equation \eqref{eq:eq_for_n_general} can be recast in a simpler form
  \begin{equation}\fl\label{eq:eq_for_n}\vec{n}''+\vec{n}=\frac{a_1\,\br{{\vec{p}_o}\br{\vec{n}\times\dsv{n}}}\vec{n}
  \times\dsv{n}+\br{a_2-a_1^2}{}m\,{\ell\,}\Omega'\dsv{n}
  }{a_1\br{m\,c+{\vec{p}_o\vec{n}}}-\br{a_2-a_1^2}{}{m\,\ell\,}{}\Omega},\qquad \Omega\equiv\abs{\dv{n}}.\end{equation}
 By taking the scalar product of both the sides  with $\dsv{n}$, we get $\Omega'=0$, provided $a_2-a_1^2\neq0$. This gives us $\dot{\Omega}=0$, that is, $\Omega(t)\equiv\Omega_o$, where $\Omega_o>0$ denotes some constant. In the exceptional case, when $a_2-a_1^2=0$, $\Omega(t)$ remains indefinite. Nonetheless, even in the latter situation, $\vec{n}$ is a definite function of $s$.

 It is a simple matter to solve equation \eqref{eq:eq_for_n}.
    When $\vec{p}_o=\vec{0}$, $\vec{n}(s)$ describes a large circle on the surface of the unit sphere. When $\vec{p}_o$ is a nonzero constant vector, $\vec{p}_o\ne\vec{0}$, we define an auxiliary function $Y$ and a constant parameter $\mu$ as follows: $$Y(s)=\frac{\vec{p}_o\vec{n}(s)}{\abs{\vec{p}_o}},\qquad \mu=\frac{a_1\abs{\vec{p}_o}}{a_1mc-\br{a_2-a_1^2}m\ell\,\Omega}.$$
  In the non-relativistic limit of concern here, $\abs{\vec{p}_o}\ll mc$ and $\Omega_o\ell\ll c$; hence $\mu$ is a small number
    (when $a_2-a_1^2=0$,
  $\mu$ is still constant, irrespective of the fact that $\Omega$ is then indefinite).
 By taking the scalar product of equation \eqref{eq:eq_for_n} with $\vec{p}_o$, we obtain the following equation for function $Y(s)$:
 $\br{1+\mu{}Y}Y''+\mu Y'^2+\br{1+2\mu Y}Y-\mu=0$. In spite of its nonlinearities, this equation can be still solved, though for
 the inverse function $s(Y)$ only  $$s(Y)=\int_Y\frac{\br{1+\mu y}\ud{y}}{\sqrt{a^2+2\mu y-\br{1-\mu^2}y^2-2\mu y^3-\mu^2y^4}},$$ where $a$ is an integration constant.

  This way the problem of motion of non-relativistic rotators can be regarded as completely solved.
   When $a_2-a_1^2\ne0$, the initial value problem has a unique solution with constant frequency $\abs{\dv{n}}=\Omega_o$. When $a_2-a_1^2=0$, the initial value problem does not have a unique solution, in which case the solution depends on a single arbitrary function of time $\Omega(t)$, only the initial value of the frequency is known.

  There is another way around to see why the frequency $\Omega(t)$ remains indefinite when $a_2-a_1^2=0$. To this end, consider an energy function defined by means of the Legendre transform
   $\mathcal{G}\br{q,\dot{q}}=\dot{q}^i{\partial{}_{\dot{q}^i}L}-L$.
 Hence, the energy function associated withe the Lagrangian \eqref{eq:lagrangian} is
\begin{equation}\label{eq:energy_rotators}\mathcal{G}=\frac{1}{2m}\br{m\dv{x}-
a_1m\ell\abs{\dv{n}}\vec{n}}^2+\frac{1}{2}m\ell^2\br{a_2-a_1^2}\dv{n}^2.
\end{equation}
For solutions, the first term can be rewritten in terms of the conserved momentum $\vec{p}=\vec{p}_o$,
$$\mathcal{G}=\frac{\vec{p}_o^2}{2m}+\frac{1}{2}m\ell^2\br{a_2-a_1^2}\Omega^2(t), \qquad \Omega(t)=\abs{\dv{n}}.$$
Since $\mathcal{G}$ is also conserved for solutions
($\dot{\mathcal{G}}=-\partial_tL\equiv0$ for
 rotators),
 the conclusion is now straightforward -- conservation of energy determines $\Omega(t)$ only when $a_2-a_1^2\ne0$.

  Before proceeding further, it is worth explaining the geometrical interpretation of the above solution
  by comparing it with that known in the fully relativistic case. The solution we presented above is written in a coordinate frame in which the center of mass moves with velocity $\frac{1}{m}\vec{p}_o$ along a straight line, however, the path on the unit sphere traced out by the director $\vec{n}(s)$ is not an exact circle, but departs slightly from it, unlike in the relativistic case.
       Had we included more terms in the relativistic expansion of the original Lagrangian, the path would be closer to a circle. Anyway, in the non-relativistic limit,  when $\mu$ is a small number, not all terms in the above integral expressing $s(Y)$ are important, and we obtain approximately $Y(s)=\mu+ a\cos\br{s-s_o}$, $a=Y(s_o)-\mu$, that is, a circle lying on a sphere. This circle can be regarded as the image of a conformally transformed large circle, formed as the result of a Lorentz boost with velocity $\frac{1}{m}\vec{p}_o$.

\section{Degenerate Hessian quadratic forms and nullifying variations}
In the previous section we came to the conclusion that the family of non-relativistic rotators defined by the Lagrangian \eqref{eq:lagrangian} consists of well-behaved dynamical systems, with the single exception when $a_2-a_1^2=0$, in which case the time evolution remains indefinite. This occurs for two rotators $a_1=\pm\sqrt{a_2}$ (provided $a_2>0$), which are non-relativistic counterparts of fundamental relativistic rotators (with $f(x)=\sqrt{1\mp\sqrt{x}}$) whose evolution is also indefinite. In this section, we shall analyze the reason for the indeterministic behavior in a way analogous to that given in \cite{bib:bratek_1}.

The  Lagrangian of non-relativistic rotators \eqref{eq:lagrangian} can be equivalently written as
$$L=\frac{1}{2}m\br{\dv{x}-\,a_1\ell\abs{\dv{n}}\vec{n}}^2+
\frac{1}{2}m\ell^2\br{a_2-a_1^2}\dv{n}^2-\,a_1m\ell\,c\abs{\dv{n}},$$ or, if $a_2\neq a_1^2$, as
$$L=\frac{1}{2}m\br{\dv{x}-\,a_1\ell\abs{\dv{n}}\vec{n}}^2+
\frac{1}{2}m\ell^2\br{a_2-a_1^2}\br{\abs{\dv{n}}-\frac{c}{\ell}
\frac{a_1}{a_2-a_1^2}}^2-\frac{1}{2}mc^2\frac{a_1^2}{a_2-a_1^2}.$$
In order to find out whether an extremal of the action functional evaluated for the Lagrangian could furnish a local minimum  with respect to comparison functions satisfying given boundary conditions, it is necessary to examine the second variation of the functional (which we shall call the Hessian (quadratic) form).  The (strong) Legendre necessary criterion states  that this happens  when the Hessian  form is positive definite \cite{bib:hilbert} (the first variation vanishes on account of Euler-Lagrange equations).
The Hessian form associated with the Lagrangian \eqref{eq:lagrangian} is\myfootnote{One can verify this by finding the $\mathcal{O}\br{\eps^2}$ term in a Taylor series expansion of the Lagrangian evaluated at comparison functions $\vec{x}+\eps\delta\vec{x}$ and $\vec{n}+\eps\delta\vec{n}$, and preserving only the terms quadratic in the variations of velocities. The other method is to note first that $\frac{\partial^2L}{\partial{\dot{q}^i}\partial{\dot{q}^j}}\delta\dot{q}^i\delta\dot{q}^j
=\frac{\partial^2L}{\partial{\dv{x}}\partial{\dv{x}}}\delta\dv{x}\delta\dv{x}
+2\frac{\partial^2L}{\partial{\dot{u}^a}\partial{\dv{x}}}\delta\dot{u}^a\delta\dv{x}
+\frac{\partial^2L}{\partial{\dot{u}^a}\partial{\dot{u}^b}}\delta\dot{u}^a\delta\dot{u}^b$, and next that $\frac{\partial^2L}{\partial{\dot{u}^a}\partial{\dv{x}}}\delta\dot{u}^a\delta\dv{x}=
\frac{\partial^2L}{\partial{\dv{n}}\partial{\dv{x}}}\delta\dv{n}\delta\dv{x}$, since $\frac{\partial^2L}{\partial{\dot{u}^a}\partial{\dv{x}}}=
\frac{\partial^2L}{\partial{\dot{n}}\partial{\dv{x}}}
\frac{\partial{\dv{n}}}{\partial{\dv{u}^a}}$
and
$\frac{\partial^2L}{\partial{\dot{u}^a}\partial{\dot{u}^b}}=
\frac{\partial}{\partial{\dot{u}^a}}\br{\frac{\partial{}L}{
\partial{\dv{n}}}\frac{\partial{\dv{n}}}{\partial{\dot{u}^b}}}=
\frac{\partial{\vec{n}}}{\partial{{u}^b}}\frac{\partial}{\partial{\dot{u}^a}}\br{\frac{\partial{}L}{\partial{\dv{n}}}}=
\frac{\partial{\vec{n}}}{\partial{{u}^a}}\frac{\partial{\vec{n}}}{\partial{{u}^b}}{\frac{\partial^2L}{\partial\dv{n}\partial\dv{n}}}=
\frac{\partial{\dv{n}}}{\partial{\dot{u}^a}}
\frac{\partial{\dv{n}}}{\partial{\dot{u}^b}}{
\frac{\partial^2L}{\partial\dv{n}\partial\dv{n}}}$.
Here, $u^a$ denote the internal coordinates on the unit sphere $\vec{n}^2=1$.}
\begin{eqnarray}\fl \nonumber \frac{1}{2}\frac{\partial^2L}{\partial{\dot{q}^i}\partial{\dot{q}^j}}
\delta\dot{q}^i\delta\dot{q}^j=
\frac{1}{2}m\br{\delta\dv{x}-a_1\ell\frac{
\dv{n}\delta\dv{n}}{\abs{\dv{n}}}\vec{n}}^2
+\frac{1}{2}m\ell^2\br{a_2-a_1^2}\br{\frac{\dv{n}\delta\dv{n}}{\abs{\dv{n}}}}^2
\\\label{eq:Hessian_form}+\frac{1}{2}m\ell c\frac{\br{-a_1}}{\abs{\dv{n}}}
\br{1+\frac{\vec{n}\dv{x}}{c}-\frac{\ell}{c}\frac{a_2}{a_1}\abs{\dv{n}}}
\br{\frac{\dv{n}\times\delta\dv{n}}{\abs{\dv{n}}}}^2.\end{eqnarray}
The quadratic form \eqref{eq:Hessian_form} is positive definite when $a_1<0$ and $a_2-a_1^2>0$ (in the non-relativistic limit, the expression  ${1+\frac{\vec{n}\dv{x}}{c}-
\frac{\ell}{c}\frac{a_2}{a_1}\abs{\dv{n}}}$ is always positive).
Rotators with $a_2-a_1^2<0$ violate the Legendre necessary criterion (however, the solutions  are still stationary points, or more precisely, saddle points of the action functional, if the second variation of the functional is studied solely in terms of the properties of the Hessian form).

The critical case, when $a_2-a_1^2=0$, requires particular attention and is central for the understanding of the dynamics and interactions with external fields.
 Although the action might still attain a minimum in this case, this cannot be detected by the Legendre necessary criterion.
 For our needs, however, this  analysis is sufficient because, as we shall see, already the dynamical systems with degenerate Hessian quadratic forms exhibit sufficiently pathological behavior not to be considered as physically viable models of particles.

\subsection{Degenerate Hessian forms}

In this section we shall be  studying the case of a degenerate Hessian form. The determinant of the Hessian matrix associated with this form, called for short the Hessian determinant, vanishes identically.
A degenerate Hessian form  behaves as a quadratic form with a lower number of degrees of freedom and the associated Hessian matrix has a nontrivial kernel.
In general, when $\mathcal{H}$ is the Hessian matrix associated with a Lagrangian $L$ and $\det{\mathcal{H}}=0$, there exists at least one nonzero vector $\eta$ such that $\mathcal{H}\eta=0$ and $\eta^T\mathcal{H}=0$ ($\mathcal{H}=\mathcal{H}^T$). In this case, the dynamical system is called defective.

In what follows, we specialize our considerations to the particular example of Lagrangian \eqref{eq:lagrangian}.
First suppose that a variation $\delta\dv{n}$ is parallel to $\dv{n}$,  $\dv{n}\times\delta\dv{n}\equiv0$.
In this case, we can separate the quadratic term associated with the degree of freedom that is responsible for variations in the direction perpendicular to $\dv{n}$. This degree of freedom is absent from the first two terms in \eqref{eq:Hessian_form} (note that  the variation $\delta\dv{n}$ has two degrees of freedom since, similarly as vector $\dv{n}$, it must lie in the plane perpendicular to the unit vector $\vec{n}$). In effect, we obtain a cutoff quadratic form for the other four degrees of freedom, three ones associated with the variations of  $\dv{x}$ and the fourth one associated with the variations in the magnitude of $\dv{n}$. Up to an unimportant factor, the cutoff quadratic form reads
$${\vec{v}^2-2u\br{\vec{n}\vec{v}}+\frac{a_2}{a_1^2}u^2},\qquad \vec{v}=\delta\dv{x},\quad {u}=a_1\ell\frac{\dv{n}\delta\dv{n}}{\abs{\dv{n}}}.$$
This form is non-degenerate only when $a_2-a_1^2\ne0$, otherwise it is degenerate. Indeed, when $a_2-a_1^2=0$, it can be brought into the canonical form
$\vec{w}\vec{w}$ with matrix $\mathrm{diag}[1,1,1,0]$, where $\vec{w}=\vec{v}-u\vec{n}$. Consequently, for $a_2-a_1^2=0$, the cutoff quadratic form is nonnegative and vanishes for $\vec{v}=u\vec{n}$. The
$4\times4$ square matrix associated with this form has a vanishing determinant and its rank is $3$.
These considerations can now be extended, so as to cover the case of the original Hessian form \eqref{eq:Hessian_form}. This form is degenerate only when $a_2-a_1^2=0$, in which case the determinant of the associated $5\times5$ Hessian matrix vanishes. In a coordinate notation, the condition $\dv{n}\times\delta\dv{n}=0$, we assumed above, is equivalent to $\dot{\theta}\delta\dot{\phi}=\dot{\phi}\delta\dot{\theta}$, which implies that
$\delta\dot{\phi}=\mathcal{A}\dot{\phi}$ and $\delta\dot{\theta}=\mathcal{A}\dot{\theta}$, with $\mathcal{A}$ being some function. Furthermore, ${\dv{n}}\delta{\dv{n}}=\dot{\theta}
\delta\dot{\theta}+\sin^2{\theta}\dot{\phi}\delta\dot{\phi}=
\mathcal{A}\abs{\dv{n}}^2$, which gives $\delta{\dv{x}}=
a_1\ell\mathcal{A}\vec{n}\abs{\dv{n}}$ when $\vec{w}=\vec{0}$. In a vector notation, $\delta{\dv{n}}=\delta({\dot{\theta}\partial_{\theta}\vec{n}+
\dot{\phi}\partial_{\phi}\vec{n}})={\partial_{\theta}\vec{n}\,\delta\dot{\theta}+
\partial_{\phi}\vec{n}}\,\delta\dot{\phi}$, hence the variation of $\dv{n}$ corresponding to $\vec{w}=\vec{0}$ can be re-expressed in the form $\delta{\dv{n}}=\mathcal{A}\dv{n}$. Finally, up to the unimportant functional factor, the variation  can be written as
\begin{equation}\label{eq:null_variation}\delta\dv{x}=\varepsilon{}a_1\ell\abs{\dv{n}}\vec{n},\qquad
\delta{\dv{n}}=\varepsilon\dv{n},
\qquad\abs{\varepsilon}\ll1.\end{equation} Here, $\varepsilon$ indicates that the variation is small. When $a_2-a_1^2=0$, we call the variation a \textit{nullifying variation}, since then it is associated with the degeneracy of the Hessian form.
When, $a_2-a_1^2=0$, the Hessian form is nonnegative if $a_1<0$, however, it attains zero for a nonzero variation \eqref{eq:null_variation}, whereas a nonnegative non-singular quadratic form could be zero only for  $\delta\dv{x}=\vec{0}$ and $\delta\dv{n}=\vec{0}$. We see once more that the Hessian form is degenerate when $a_2-a_1^2=0$ (we exclude the trivial case $a_1=0$ and $a_2=0$ of a point particle with three degrees of freedom).

The existence of a nontrivial nullifying variation is  associated  with the vanishing of the  Hessian determinant.
The Hessian determinant corresponding to the velocities $\dot{q}=(\dv{x}, \dot{\theta},\dot{\phi})$ reads
$$\mathrm{Det}\left[\frac{\partial^2L}{
\partial{\dot{q}^{i}}\partial{\dot{q}^{j}}}\right]
=m^{4}\ell^{2}\br{a_2-a_1^2}\frac{m\ell^2}{\frac{\ell}{c}\abs{\dv{n}}}
\br{a_2\frac{\ell}{c}\abs{\dv{n}}-{}a_1 \br{1+\frac{\vec{n}\vec{\dv{x}}}{c}}}\sin^2{\theta},$$
or,
using the definition of momentum $\vec{\pi}$,  \begin{equation}\label{eq:determinant} \mathrm{Det}\left[\frac{\partial^2L}{
\partial{\dot{q}^{i}}\partial{\dot{q}^{j}}}\right]
=m^{4}\ell^{2}\br{a_2-a_1^2}\frac{\vec{\pi}\dv{n}}{\dv{n}\dv{n}}
\sin^2{\theta}.\end{equation}
This determinant vanishes identically when $a_2-a_1^2=0$, as expected. This fact implies that both columns and rows of the (symmetric) Hessian matrix are linearly dependent when $a_2-a_1^2=0$. As so, there is a  nontrivial linear combination of columns or rows of the matrix which equals zero. Actually, up to normalization, there is only a single vector of coefficients of this combination, since, as we have seen from the above Hessian form analysis, the Hessian matrix has rank $4$. It therefore has a one-dimensional kernel, called the null space of the Hessian matrix and spanned by this vector. We call this vector a \textit{nullifying vector} of the Hessian and denote it by $\eta$.

The nullifying vector is directly related to the nullifying variation $\delta\dv{x}=\varepsilon{}a_1\ell\abs{\dv{n}}$, $\delta\dot{\theta}=\varepsilon\dot{\theta}$ and $\delta\dot{\phi}=\varepsilon\dot{\phi}$, we have already encountered. Indeed, by a direct calculation, one can verify that the nullifying vector is proportional to this variation, that is, $\eta=a_1\ell\abs{\dv{n}}\vec{n}\oplus[\dot{\phi},\dot{\theta}]$. Represented as a column vector, the nullifying vector reads
\begin{equation}
\label{eq:nullifying_vector}
\eta=[a_1\ell\abs{\dv{n}}n_x,a_1\ell\abs{\dv{n}}n_y,a_1\ell\abs{\dv{n}}n_z,
\dot{\phi},\dot{\theta}]^T.
\end{equation}
This vector is also determined up to an unimportant normalization factor.

 \section{Hessian singularity and integrability}\label{sec:integrability}
It would be interesting to
see the relationship between uniqueness of the initial value problem and the properties of the Hessian matrix.
 The equations of motion of a mechanical system defined by a Lagrangian $L$ in expanded form read
 \begin{equation}\label{eq:ELequations_expanded}\ddot{q}^j\frac{\partial^2L}{\partial{\dot{q}^j}\partial{\dot{q}}^i}
=-
\dot{q}^j\frac{\partial^2L}{\partial{q^j}\partial{\dot{q}}^i}-
\frac{\partial^2L}{\partial{t}\partial{\dot{q}}^i}
+\frac{\partial L}{\partial q^i}.\end{equation}
As is well known, solutions to these equations coincide with the extremals of the Hamilton action and provide the law of motion for a mechanical system.
Let us now assume that the Hessian matrix on the right-hand side of equation \eqref{eq:ELequations_expanded} is non-singular. Then, at any instant of time, given actual positions and velocities,    there is only a single acceleration vector possible. The acceleration vector can be determined algebraically by solving the above equations for $\ddv{q}$. This can be done, since a non-singular Hessian matrix is invertible. However, when the Hessian matrix is singular, it is not invertible and, consequently, the accelerations cannot be uniquely determined from the actual positions and velocities, at any instant of time. In other words, given positions and velocities, there are infinite number of accelerations available, among which
a defective system could choose at each stage of its motion.

We have already seen a system with an indefinite acceleration vector. The frequency of rotation $\Omega(t)=\abs{\dv{n}}$ has been found to be indefinite for the
rotator whose Hessian determinant \eqref{eq:determinant} vanishes identically.
It is easy to understand the mechanism of why this arbitrary function is present in the solution. Suppose that, at some instant of time, a defective dynamical system  spontaneously changes its accelerations in the direction of the nullifying vector.
The reason for this change is unimportant and reflects only an instability pertinent in systems with singular Hessian.
The system can be perturbed in this way with no energy cost. To see this, consider
the energy function $\mathcal{G}$ defined by the Legendre transform of the Lagrangian $L$,
$\mathcal{G}\br{q,\dot{q}}=\dot{q}^i{\partial{}_{\dot{q}^i}L}-L$.
A variation of $\mathcal{G}$ (without the variation of time) reads
$$\delta\mathcal{G}=\dot{q}^i\frac{\partial^2L}{\partial\dot{q}^i\partial\dot{q}^k}\delta
\dot{q}^k+\br{\dot{q}^i\frac{\partial^2L}{\partial\dot{q}^i\partial{q}^k}-
\frac{\partial{}L}{\partial{q}^k} }\delta{q}^k.$$
Suppose now, that at some instant of time, the velocities have been changed in the direction of a nullifying vector $\delta{\dot{q}}^k=\varepsilon\eta^k$ without altering the position of the system, then
  $\delta\mathcal{G}=0$. Null variations $\delta\dv{x}=a_1\ell\varepsilon\abs{\dv{n}}\vec{n}$ and $\delta{\dv{n}}=\varepsilon\dv{n}$ are responsible for the change in the frequency of rotation $\Omega$
  of a defective rotator. Indeed, if $\Omega=\abs{\dv{n}}$, then $\dv{n}\delta\dv{n}=\Omega\delta\Omega$ and, since $\dv{n}\times\delta\dv{n}=0$, $\delta{\Omega}=\Omega^{-1}\abs{\dv{n}}\abs{\delta\dv{n}}=\varepsilon\Omega$, which gives $\delta\dv{x}=a_1\ell\vec{n}\delta{\Omega}$, compare with \eqref{eq:null_variation}. This means that spinning a defective rotator up or down costs no energy.

Note that when the Hessian matrix is degenerate, the principle of the least action does not discriminate between various solutions satisfying the same initial conditions. In this case, various comparison motions can be indicated which do not change the value of the action functional, in which case the comparison motions provide other possible solutions. Exactly this happens for the defective rotator -- the action value taken between two fixed instants of time remains the same when the frequency of rotation is changed within that time interval.

One could also release the requirement of the least action principle, and instead, apply the principle of stationary action. But again, it is indispensable that the Hessian form (although indefinite in general) be non-degenerate, in order to ensure uniqueness of accelerations. Based on the least action principle,
we would have to assume that $a_1<0$ and $a_2>a_1^2$,  to make the Hessian form \eqref{eq:Hessian_form} strictly positive, whereas, based on the stationary action principle, it would suffice to assume that $a_2-a_1^2\ne0$ only.

\subsection{The relationship with the Hamiltonian formulation}
Suppose that the Lagrangian of a mechanical system is expressed in some coordinates $q$ compatible with the constraints and that the corresponding generalized velocities are $\dot{q}$ (we remind that $q$ denotes the dynamical degrees of freedom, and all auxiliary or gauge degrees of freedom are assumed to have been already eliminated). The canonical momenta conjugate to the velocities are
$p_i=\partial_{\dot{q}^i}L$.
From the inverse function theorem it follows that,  given positions,
the velocities can be regarded as functions of momenta, $\dot{q}=v(q,p)$, provided that the system of differentials $dp_i=\mathcal{H}_{ij}\ud{\dot{q}}^{j}$ is invertible, where again the Hessian matrix has appeared
$$\mathcal{H}_{ij}=\frac{\partial^2L}{\partial{\dot{q}^i}\partial{\dot{q}^j}}.$$
Only in this case the energy function $\mathcal{G}(q,\dot{q})$, defined by means of the Legendre transform, gives rise to a Hamiltonian $H(q,p)=\mathcal{G}(q,v(q,p))$.
 It is thus indispensable for the existence of the Hamiltonian formulation that the Hessian determinant be nonzero, otherwise the map between velocities and momenta would not be invertible, and consequently, the Hamiltonian and Lagrangian pictures would not be equivalent.

 By applying this textbook knowledge to rotators, we conclude that
 if $a_2-a_1^2\neq0$, the velocities $\dv{x}$ and $\dv{n}$ can be expressed locally by the conjugate canonical momenta. This is done by solving the equations
defining the momenta \eqref{eq:momenta} (given a direction vector $\vec{n}$, only two components of $\vec{\pi}$ are  independent, since $\vec{\pi}\vec{n}\equiv0$).
On expressing $\vec{n}\dv{x}$ in the definition of $\vec{\pi}$  by $\vec{p}\vec{n}$, found from the definition of $\vec{p}$ in equation \eqref{eq:momenta}, and calculating $\abs{\vec{\pi}}$, one arrives at the following equation for $\abs{\dv{n}}$:
\[\abs{\dv{n}}=\frac{c}{\ell}\frac{ a_1}{a_2-a_1^2}\br{1+\frac{\vec{p}\vec{n}}{m c}-\frac{1}{\abs{a_1}}\frac{\abs{\vec{\pi}}}{m\ell c}}.\] It follows that the sign of the expression in the bracket must be the same as the sign of $\frac{a_1}{a_2-a_1^2}$. This remark is useful for deriving the   Hamiltonian.
 The velocities expressed by positions and momenta read
 \begin{eqnarray*}
\frac{\dv{x}}{c}=\frac{\vec{p}}{m c}+\frac{a_1^2}{a_2-a_1^2}\br{1+\frac{\vec{p}\vec{n}}{m c}-\frac{1}{\abs{a_1}}\frac{\abs{\vec{\pi}}}{m \ell c}}\vec{n},\\
\frac{\ell}{c}\dv{n}=-\frac{\abs{a_1}}{a_2-a_1^2}\br{1+\frac{\vec{p}\vec{n}}{m c}-\frac{1}{\abs{a_1}}\frac{\abs{\vec{\pi}}}{m \ell c}}\frac{\vec{\pi}}{\abs{\vec{\pi}}}.\end{eqnarray*}
The Hamilton function for all non-degenerate rotators can now be found by means of the Legendre transform,
$$H=\frac{\partial L}{\partial \dv{x}}\dv{x}+\frac{\partial L}{\partial \dv{n}}\dv{n}-L,$$
where we have used the following implication:
$$\frac{\partial\dv{n}}{\partial\dot{\theta}}=\frac{\partial\vec{n}}{
\partial{\theta}},\quad \frac{\partial\dv{n}}{\partial\dot{\phi}}=\frac{\partial\vec{n}}{
\partial{\phi}},\quad\Rightarrow\quad\frac{\partial L}{\partial \dot{\theta}}\dot{\theta}+\frac{\partial L}{\partial \dot{\phi}}\dot{\phi}=\frac{\partial L}{\partial \dv{n}}\dv{n}.$$
Hence, the Hamilton function reads
$$H=\frac{\vec{p}^2}{2m}+\frac{m c^2}{2}\frac{a_1^2}{a_2-a_1^2}\br{1+\frac{\vec{n}\vec{p}}{m c}-\frac{1}{\abs{a_1}}\frac{\abs{\vec{\pi}}}{m\ell c}}^2,\qquad \vec{n}\vec{\pi}=0, \quad\vec{n}\vec{n}=1.$$
In the above derivation, it was assumed that the sign of the expression $ {a_2\frac{\ell}{c}\abs{\dv{n}}-{}a_1 \br{1+\frac{\vec{n}\vec{\dv{x}}}{c}}}$ is the same as that of  $\br{-{}a_1}$. This is clearly satisfied in the non-relativistic limit.

The map between velocities and momenta becomes singular when $a_2-a_1^2=0$, in which case the Hamiltonian is not defined (although the energy function  can be still defined).
If, nevertheless, one would insist upon ignoring the singularity of the Hessian matrix (or simply overlooked this singularity), the Hamiltonian picture might lead to a dynamics inequivalent to that suggested by the Lagrangian picture (possibly overlapping for special motions). For example, by using the definition of momentum $\vec{p}$ in \eqref{eq:momenta}, the energy function of a degenerate rotator $\mathcal{G}
=\frac{1}{2}m\br{\dv{x}-a_1\ell\abs{\dv{n}}\vec{n}}^2$ (compare with \eqref{eq:energy_rotators}), can be formally rewritten as a false Hamiltonian, $"H"=\frac{\vec{p}^2}{2m}$, which, as the only solution, gives the motion of the center of mass (with the  other degrees of freedom fixed).

\medskip

As we have seen, indispensable for a well-defined dynamics of a mechanical system described by the dynamical degrees of freedom only is the non-singularity of the Hessian matrix; otherwise,
the motion of the system might become indefinite, that is, not uniquely determined from the stationary action principle and from the initial conditions.
 One could surmise that by adding to the Lagrangian an interaction term with an external field, the situation could be remedied. Right, it would happen when this interaction could render the Hessian non-singular, but this would not solve the central problem of the free motion, at all. Even if one has ignored the problem of indefiniteness of the free motion, the fact that the additive interaction term could not be linear in velocities cannot be ignored,  since such a form of interaction would not change the already singular Hessian matrix even in the presence of external fields. This remark excludes, in particular, the ordinary, minimal interaction with the external electromagnetic field. We shall illustrate in the next section that even when such interaction turns out to make a particular solution definite, this happens only occasionally, by a special choice of the field configuration and initial conditions.

\subsection{\label{sec:linearity}Hessian  singularity and constraints}

The degeneracy of the Hessian matrix addressed in the preceding sections, implies the existence of some first order integrability conditions that must be imposed on the solutions of the equations of motion.
Let us suppose that the Hessian matrix of a system described by a Lagrangian $L$, is singular. Then, the Hessian matrix possess a nontrivial nullifying vector $\eta$ (at least one) such that
$$a^j\frac{\partial^2L}{\partial{\dot{q}^j}\partial{\dot{q}}^i}\eta^i\equiv0,$$
for any vector $a^i$.
As so, we obtain from the equations of motion \eqref{eq:ELequations_expanded}, the following differential constraint of the first order
\begin{equation}\label{eq:constraint_general}\br{
\dot{q}^j\frac{\partial^2L}{\partial{q^j}\partial{\dot{q}}^i}+
\frac{\partial^2L}{\partial{t}\partial{\dot{q}}^i}
-\frac{\partial L}{\partial q^i}}\eta^i=0.\end{equation} This constraint is trivially satisfied for a well-behaved  dynamical system, that is, that with non-singular Hessian matrix, in which case the kernel of the Hessian matrix is trivial, $\eta^{i}\equiv0$.
However,  the constraint will usually turn out nontrivial for defective dynamical systems. Interestingly, this constraint is still trivial for defective rotators, despite  the fact that the nullifying vector is then nonzero. Indeed, one can prove  the following identity for the Lagrangian \eqref{eq:lagrangian}
$$a_2=a_1^2\qquad\Rightarrow\qquad\br{\frac{\ud{}}{\ud{t}}\frac{\partial L}{\partial \dot{q}^i}-\frac{\partial L}{\partial q^i}}\eta^i\equiv0,$$ where $\eta$ is given in \eqref{eq:nullifying_vector}.
This identity holds for defective rotators irrespectively of whether the equations of motion are satisfied or not. Hence, the equations of motion are linearly dependent  when $a_2=a_1^2$, which is the reason for the presence of
     arbitrary function of time in the general solution. Constraint \eqref{eq:constraint_general} may turn out nontrivial when an additional interaction term is added to the Lagrangian \eqref{eq:lagrangian} with $a_2=a_1^2$, such that the Hessian of the new Lagrangian remains singular.
     Could this constraint make motion of a defective dynamical system unique in spite of the Hessian singularity?

\section{Motion of a defective dynamical system in the electromagnetic field}

As we have noted in the previous section, the example of an interaction which does not remove the Hessian singularity, is the ordinary interaction with the electromagnetic field, commonly assumed for electrically charged structure-less point particles.
       The Lagrangian of a non-relativistic rotator, treated as a structure-less (point) particle with electric charge $e$,
in the presence of the external electromagnetic field, is
$$L_{EM}=\frac{1}{2}m\dot{\vec{x}}^2+\frac{1}{2}a_2m\ell^2\dot{\vec{n}}^2
- a_1 m \ell c \abs{\dot{\vec{n}}}\br{1+ \frac{\vec{n}\dot{\vec{x}}}{c}}+L_{INT},\qquad L_{INT}=\frac{e}{c}\vec{A}\dv{x}-e\Phi.$$
In fact, a rotator is not structureless, since it has two intrinsic degrees of freedom that also might couple to the electromagnetic field (through the intrinsic magnetic moment, etc).
It is evident that
$$\frac{\partial^2L_{INT}}{\partial{\dot{q}^j}\partial{\dot{q}}^i}=0,$$
hence the Lagrangian $L_{EM}$ has the same nullifying vector as the Lagrangian of a free rotator \eqref{eq:lagrangian} (we remind that, when $a_2-a_1^2\ne0$, the nullifying vector is trivial zero, when $a_2-a_1^2=0$, it is nontrivial and given by equation \eqref{eq:nullifying_vector}).
As we have seen in section \ref{sec:linearity}, for a defective electrically charged rotator in free motion, the constraint \eqref{eq:constraint_general} happens to be satisfied identically, that is, not only for solutions. This means that the equations of motion are not independent (the same holds for electrically charged counterparts of both  fundamental relativistic rotators \cite{bib:bratek_4}).
In the presence of the external electromagnetic field, we thus obtain for a defective rotator the following constraint (with $\eta$ given in equation \eqref{eq:nullifying_vector})
$$\br{
\dot{q}^j\frac{\partial^2L_{INT}}{\partial{q^j}\partial{\dot{q}}^i}+
\frac{\partial^2L_{INT}}{\partial{t}\partial{\dot{q}}^i}
-\frac{\partial L_{INT}}{\partial q^i}}\eta^i=0$$
On introducing the customary definitions of the electric field $\vec{E}\equiv-\frac{1}{c}\partial_t\vec{A}-\vec{\nabla}\Phi$ and of the magnetic field $\vec{H}\equiv\vec{\nabla}\times\vec{A}$, after simple calculation, the above condition can be recast in a more transparent form
\begin{equation}\label{eq:constraint_EM}
\vec{n}\circ\br{\frac{\dv{x}}{c}\times\vec{H}+\vec{E}}=0.\end{equation}
This constraint says that for consistency with the Hessian singularity of defective rotators, the direction vector must be always perpendicular  to the Lorentz force.
Constraint \eqref{eq:constraint_EM} is nothing but an integrability condition for the equations of motion of a degenerate rotator in the electromagnetic field. This constraint must be satisfied in the course of the rotator's motion. In particular, the constraint must be satisfied by the initial conditions. As an aside, it should be remarked that the relativistic counterpart of this condition, derived for an electrically charged (defective) fundamental relativistic rotator, is slightly different, $\vec{n}\circ\br{\frac{\dv{x}}{c}\times\vec{H}+\vec{E}}
=\frac{\dv{x}}{c}\circ\vec{E}$, which can be written as $F_{\mu\nu}k^{\mu}\dot{x}^{\nu}=0$, compare with \cite{bib:bratek_4}). We stress again that such a constraint cannot appear when the Hessian matrix of a free dynamical system is non-singular (compare with the  further discussion).

There is another way of showing that constraint \eqref{eq:constraint_EM} must hold for a degenerate rotator. It is conceptually unrelated to the previous derivation of the constraint.
To this end, consider the energy function of a non-relativistic rotator in the electromagnetic field
$$\mathcal{G}_{EM}(q,\dot{q},t)\equiv\frac{\partial L_{EM}}{\partial{\dot{\vec{q}}}}\dot{\vec{q}}-L_{EM}=\frac{1}{2m}\br{m\dv{x}
-a_1m\ell\abs{\dv{n}}\vec{n}}^2+\frac{1}{2}m\ell^2\br{a_2-a_1^2}\dv{n}^2+e\Phi.$$
Since during the motion $\dot{\mathcal{G}}_{EM}=-\partial_tL_{EM}$ and $\dv{p}=e\br{\frac{\dv{x}}{c}\times\vec{H}+\vec{E}}$ (where $\vec{p}$ has been defined in equation \eqref{eq:momenta}), we get
$$a_1e\ell\abs{\dv{n}}\vec{n}\circ\br{\frac{\dv{x}}{c}\times\vec{H}+\vec{E}}=
m\ell^2\br{a_2-a_1^2}\abs{\dv{n}}\frac{\ud{}\abs{\dv{n}}}{\ud{t}}.$$
When the rotator is uncharged and $\abs{\dv{n}}\ne0$, this means that the frequency of rotation must be constant, provided $a_2-a_1^2\neq0$. In general, when $a_2-a_1^2\neq0$, the above formula
gives us the evolution law for the frequency of rotation
$$\dot{\Omega}=\frac{a_1}{a_2-a_1^2}\frac{e}{m\ell}\,\vec{n}
\circ\br{\frac{\dv{x}}{c}\times\vec{H}+\vec{E}}, \qquad a_2-a_1^2\ne0.$$
In the absence of the electromagnetic field (or when $e=0$), the frequency is constant and the initial conditions can be arbitrary, as expected for a well-behaved dynamical system.
However, in the exceptional case of a defective rotator, when $a_2=a_1^2$, the energy conservation
   does not give us any law of frequency evolution. Instead, one again is led to constraint \eqref{eq:constraint_EM}, which, starting from quite different premises, has been  already predicted for electrically  charged defective  rotators.  When $a_2=a_1^2$, the direction vector $\vec{n}$ cannot be oriented arbitrarily, but  must be always perpendicular to the Lorentz force.

\medskip
There is a qualitative difference between the following two situations,
despite the fact that in both situations we have dynamical systems constructed in the same way.
 In the first situation, we consider the motion of an electrically charged defective rotator and, in the second situation, we consider the motion of all other rotators, that is, electrically charged or electrically neutral rotators with non-singular Hessian
  and electrically neutral defective rotator.
 While in the second situation all initial conditions are treated on equal footing (irrespectively of Hessian singularity), as it should be for dynamical systems with a definite number of degrees of freedom, there is a constraint limiting this freedom in the first situation. It is clear that this limitation in the freedom of choosing the initial data has nothing to do with the presence of the electric charge or the electromagnetic field.
 From the physical point of view, such a limitation should be considered as a defect. Putting this differently, the form of the interaction with the electromagnetic field assumed for a defective rotator is incompatible with the Hessian singularity.

Constraint \eqref{eq:constraint_EM} leads to various paradoxes. For example, on the one hand, in the presence of the electromagnetic field,
an electrically charged defective rotator cannot start its motion from the same initial conditions as its electrically neutral counterpart, while, on the other hand, the electrically neutral defective rotator can always start from the identical initial conditions like an electrically charged or electrically neutral rotator with non-degenerate Hessian. It is evident from this paradox  that the presence of constraint \eqref{eq:constraint_EM} has nothing to do with the form of the interaction with the electromagnetic field assumed in the Lagrangian. The constraint appears only because the Hessian is singular. When the Hessian is non-singular,
the motion of electrically charged rotators  in the same field and with the same interaction term is not constrained  (then also no constraints are imposed on the initial conditions).
One can multiply many other suggestive examples. Consider,
a very massive defective rotator of arbitrarily small electric charge,
initially rotating very slowly. In order to satisfy constraint $\vec{n}\vec{E}=0$, to which \eqref{eq:constraint_EM} reduces
in the homogenous electric field, the rotator could not
initially point  in any direction not orthogonal to $\vec{E}$, irrespective of how weak could be the field. As another example, consider the rotator in the electromagnetic wave of arbitrarily large frequency and low intensity. Then to satisfy constraint \eqref{eq:constraint_EM}, the rotator would have to instantaneously follow any changes in this field.
One should agree that considering a defective dynamical system as a physically viable geometrical model of a spinning particle would be physically unjustified.
   Introduction of the electric charge to a defective system could be justified only when this was made in a way that could enable one to structurally change the system, rendering its Hessian form always non-degenerate.

A word of
warning may be appropriate here. Below is given a simple example which illustrates the fact that in very particular situations, a defective system can
perform misleadingly well, resembling in its behavior a well-behaved dynamical system.  Constraint \eqref{eq:constraint_EM} imposed on the initial data may occasionally result in a unique solution for the motion of an electrically charged defective system, but this occurs only accidentally, owing to the high symmetry of a particular problem.
Before coming to this issue more explicitly, first we give another simple example in order to make it, first of all, clear that, in general, the  motion of an electrically charged defective rotator remains indefinite, despite the fact that constraint \eqref{eq:constraint_EM} has been imposed, and also that, by no means, the inclusion of the ordinary electromagnetic interaction term in the Lagrangian of a defective rotator improves something with regard to the central deficiency associated with the Hessian singularity.

\subsection{Example I}
A constraint of the kind that is given in equation \eqref{eq:constraint_general}, or \eqref{eq:constraint_EM} in particular,
which exists merely because the Hessian is singular, by no means removes the central deficiency of a defective dynamical system. This should be stressed, since the presence of an additional constraint imposed on solutions, is likely to make the dynamics "more definite". When this happens for a defective dynamical system, it may be misleading, since one, after inclusion of an interaction term and finding
a unique solution, might be led to the wrong conclusion that the interaction (which, in fact, did not remove the Hessian singularity) made the system  well-behaved.

 In order to see this, consider an electrically  charged defective rotator,
moving in the uniform electric field $\vec{E}$. Constraint \eqref{eq:constraint_EM} reduces in this case to a holonomic constraint $\vec{E}\vec{n}=0$, therefore, it can be satisfied automatically by choosing appropriate generalized coordinates compatible with this constraint (in the relativistic case we would be led to a constraint $\vec{E}\br{\vec{n}-\vec{v}/c}=0$, or in coordinates, $z(t)=\int\cos{\theta(t)}\ud{t}$, much more difficult to tackle with).  This gives us  the following Lagrangian
$$L_E=\frac{1}{2m}\br{\dot{x}^2+\dot{y}^2+\dot{z}^2+a_1^2\ell^2\dot{\psi}^2}-a_1 m\ell c\abs{\dot{\psi}}\br{1+\frac{\dot{x}}{c}\cos\psi+
\frac{\dot{y}}{c}\sin\psi}+e\abs{\vec{E}}z,$$ which has been obtained from Lagrangian $L_{EM}$, assuming $a_2-a_1^2=0$.
Here, we used a coordinate system in which the $z$-axis is directed along $\vec{E}$. In this frame
$\vec{n}=\sq{\cos{\psi},\sin{\psi},0}$.
The determinant of a $4\times4$ Hessian matrix associated with Lagrangian $L_E$ is zero, we therefore expect the rotator's motion in this field to be still indefinite, in spite of the fact that the constraint \eqref{eq:constraint_EM} has been satisfied.
The nullifying vector of the reduced Hessian  is
$\eta=\{a_1\ell\eps\cos\psi,\,a_1\ell\eps\sin\psi,\,0,\,1\}^T$,
$\eps=\sgn({{\dot{\psi}}})$. Constraint \eqref{eq:constraint_EM} is
indeed satisfied,
$\br{\frac{\ud}{\ud{t}}\br{\frac{\partial{L_E}}{\partial{\dot{q}^i}}}-
\frac{\partial{L_E}}{\partial{q^i}}}{\eta^i}\equiv0.$
One can check by substitution that the extremals of the Lagrangian $L_E$ are
\begin{eqnarray*}\vec{x}(t)=\vec{x}_o+\vec{v_o}t+\{a_1\ell\eps\sin(\psi(t)),
-a_1\ell\eps\cos(\psi(t)),
\frac{e\abs{\vec{E}}}{m}t^2\}, \\ \vec{n}(t)=\{\cos(\psi(t)),\sin(\psi(t)),0\}.\end{eqnarray*} The solution contains an arbitrary function of time $\psi(t)$. Of course, this is also a solution to the equations of motion  of the original system with five degrees of freedom. At the (adiabatic) limit $\vec{E}\to0$, we obtain indefinite motion of a defective rotator in free motion.

\subsection{Example II}
Only sometimes, when a system with singular Hessian is allowed to move on a very particular sub-manifold, its motion can regain uniqueness.
In the following example, we achieve this uniqueness by devising conditions adjusted to the constraint \eqref{eq:constraint_EM} in a way to exclude the possibility of variations in the nullifying direction of the Hessian.

Namely, consider  an electrically charged defective rotator in the uniform magnetic field $\vec{H}_o$.
  Based on the analogy with the motion of an electrically charged point particle in such a  field, we consider the motion on a cylinder of radius $R$, with the main axis directed along vector $\vec{H}_o$. When $\vec{n}$ makes a constant angle $\pi/2-\alpha$ with vector $\vec{H}_o$, and the motion along axis $z$ is inertial, then the equations of motion will be satisfied only when  the azimuthal angle of $\vec{n}$ is a linear function of time. A particular solution for  $e=-Q<0$, is
         $$\vec{x}(t)=\vec{x}_o+\{-R \sin\br{\omega t},R \cos\br{\omega t},v t\},\quad \vec{n}(t)=\{\cos{\alpha}\cos{\omega t},\cos{\alpha}\sin{\omega t},\sin{\alpha}\}$$
$$\omega=\frac{\omega_L}{ a_1\frac{\ell}{R}\cos^2\alpha+ 1},\quad \omega_L=\frac{Q \abs{\vec{H}_o}}{m c},\quad v=c\frac{\sin\alpha}{\cos2\alpha}\br{1-2\frac{R \omega}{c}{\br{1+ a_1\frac{\ell}{2R}}\cos\alpha}{}},$$
where $R>0$, and we assume that $\omega>0$ as for an ordinary particle with negative charge (this requires $R>-a_1\ell\cos{}^2\alpha$ when $a_1<0$).
There are also other similar solutions.
 At the limit when the magnetic field adiabatically vanishes, the solution is reduced to the inertial motion of the center of mass with vector $\vec{n}$ pointing in a fixed direction in space. Therefore, it is not the counterpart of the indefinite free motion, which we obtained in the limit discussed in the previous example. In this sense, one can say that constraint \eqref{eq:constraint_EM} acted here to fix the frequency of rotation.

It is left only to understand why the above solution has fixed frequency.
Assume for simplicity that we consider the motion constrained to the plane orthogonal to $\vec{H}_o$ ($\alpha=0$). Then we can use a map $\vec{x}=\{-r\sin{\psi},r\cos{\psi},0\}$ and $\vec{n}=\{\cos{\phi},\sin{\phi},0\}$. The corresponding Lagrangian is
\begin{eqnarray*}\fl L=\frac{m}{2}\br{\dot{r}^2+r^2\dot{\psi}^2+a_1^2\ell^2\dot{\phi}^2}-a_1 m \ell c|\dot{\phi}|\br{1+\frac{\dot{r}}{c}\sin\br{\phi-\psi}-
\frac{r\dot{\psi}}{c}\cos\br{\phi-\psi}}\\-\frac{|\vec{H}_o| Q}{2c}r^2\dot{\psi}.\end{eqnarray*}
One can easily verify that the Hessian determinant is still zero; therefore, the plane motion does not change qualitatively the above solution.
The Hessian constraint \eqref{eq:constraint_EM} reduces to $\dot{r}\cos\br{\phi-\psi}+r\dot{\psi}\sin\br{\phi-\psi}=0$. The most trivial solution of this constraint with nonzero frequency is obtained by assuming co-rotational motion $\psi\equiv\phi$ (this gives us the analogy of the particular solutions in uniform magnetic studied in \cite{bib:schaefer}).
  With this ansatz the Hessian constraint reduces to $\dot{r}=0$, or $r=R$, where $R$ is a constant (we shall focus on the case when $R>0$). By varying the ansatz about a solution $r=R$ and $\psi=\phi$, we obtain $\delta\psi=\delta\phi$ and, by varying the Hessian constraint, we get
$\delta\dot{r}+R\dot{\psi}\br{\delta\phi-\delta\psi}=0$, that is, $\delta\dot{r}=0$. The null vector of the Hessian is $\eta=\kappa \br{a_1\ell\dot{\phi}\sin\br{\phi-\psi}\partial_{\rho}-a_1\frac{\ell}{r}
\dot{\phi}\cos\br{\phi-\psi}\partial_{\psi}+|\dot{\phi}|\partial_{\phi}}$, where $\kappa$ is some unimportant factor. In particular, for our ansatz, $\eta=\kappa \br{-a_1{\ell}{R^{-1}}
\dot{\phi}\,\partial_{\psi}+|\dot{\phi}|\partial_{\phi}}$. On the other hand, for the variations admissible by our ansatz and the Hessian constraint, we have $\delta\dot{\rho}\partial_{\rho}+\delta\dot{\psi}\partial_{\psi}+\delta\dot{\phi}
\partial_{\phi}=\delta\dot{\phi}\br{\partial_{\psi}+
\partial_{\phi}}$, which can be collinear with $\eta$ only when $-a_1\ell {\dot{\phi}}=R |\dot{\phi}|$. By solving the equations of motion, assuming our ansatz and the Hessian constraint, we obtain two particular solutions  $\dot{\phi}={\omega_L}\br{1+\frac{a_1\ell}{R}}^{-1}$ for $R>-a_1\ell$ and $\dot{\phi}={\omega_L}\br{1-\frac{a_1\ell}{R}}^{-1}$ for $R<a_1\ell$ (assuming at the same time that $R>0$). In every case,
the condition for nullifying variations cannot be satisfied
and, therefore, the solutions cannot be disturbed in the direction of the nullifying vector without violating constraint \eqref{eq:constraint_EM} (had we assumed $R<0$, we would obtain another pair of similar solutions). This time, contrary to the previous example with the electric field, constraint \eqref{eq:constraint_EM} happened to make the particular evolution definite, uniquely determined by the initial conditions. We stress again that at the limit of vanishing field, the frequency tends to zero, and one obtains the inertial motion which is qualitatively different from the free rotating motion of defective rotator.
A similar analysis of analogous solutions in the relativistic case was presented in \cite{bib:bratek_4}.

\section{Conclusions}
In this paper we considered a non-relativistic limit of a family of relativistic rotators. This enabled us in a simpler setup to study various aspects
of their dynamics, independently of this limit.
In particular, we found that the motion of non-relativistic counterparts
of the fundamental relativistic rotator and its partner is similarly indefinite as in the relativistic case (in particular, the equations of motion for the physical degrees of freedom are not independent), whereas the motion of non-relativistic counterparts of the other relativistic rotators  is definite. In every case,  indefiniteness of the motion is associated with the vanishing of the Hessian determinant (only the physical degrees of freedom are assumed to be present in the Lagrangian), and this indefiniteness cannot be removed by any interaction term linear in the velocities. Our analysis has various implications. Dynamical systems with singular Hessian are unphysical and should not be considered as sensible models of particles.
We illustrated this statement in various ways, in particular, by considering electrically charged counterparts of the rotators.
For example, while the ordinary interaction term with the electromagnetic field (the same as for electrically charged structureless point particles), leads to a well-defined dynamics when the Hessian matrix is non-singular in the free motion, there appears an unphysical constraint on the initial conditions
(and on possible motions) when the Hessian matrix is singular, since then the Hessian matrix has a nontrivial null space.
This constraint can also be derived from the energy function by taking an appropriate limit.
Occasionally, the constraint leads to a definite dynamics (which should not astonish), but in general, the motion in the external electromagnetic field remains indefinite, essentially for the same reason as in the case without the interaction term. These findings are in accord with the results obtained in the relativistic case in \cite{bib:bratek_1} and \cite{bib:bratek_4}, and the apparent contradiction between Hessian singularity and definiteness of motion encountered in \cite{bib:schaefer} is thereby clarified. The conclusions of this paper and methods developed here can be extended to other dynamical systems. In particular, these methods were successfully used to study the \frr minimally coupled to the electromagnetic field \cite{bib:bratek_4}.

\section*{Acknowledgments}
I am grateful to Zdzis{\l}aw Golda from Astronomical Observatory of the Jagiellonian University
for interesting discussions concerning the Newtonian limit of Fundamental Relativistic Rotator.

\end{document}